\documentclass[12pt,prd,aps,amssymb,amsmath,tightenlines,showpacs]{revtex4}
\usepackage{graphicx}

\def\half{{\textstyle{\frac{1}{2}}}}
\def\cP{\mathcal P}

\def\cT{\mathcal T}
\def\bea{\begin{eqnarray}}
\def\eea{\end{eqnarray}}
\begin{document}

\topmargin=0.0cm

\title{WKB Analysis of $\cP\cT$-Symmetric Sturm-Liouville problems. II}

\author{Carl~M.~Bender${}^1$}
\email{cmb@wustl.edu}

\author{Hugh~F.~Jones${}^2$}
\email{h.f.jones@imperial.ac.uk}

\affiliation{${}^1$Department of Physics, Kings College London, Strand, London
WC2R 1LS, UK \footnote{Permanent address: Department of Physics, Washington
University, St. Louis, MO 63130, USA.} \\ ${}^2$Blackett Laboratory, Imperial
College, London SW7 2AZ, UK}

\begin{abstract}
In a previous paper it was shown that a one-turning-point WKB approximation gives an accurate picture of the spectrum of certain non-Hermitian $PT$-symmetric Hamiltonians on a finite interval with Dirichlet boundary conditions. Potentials to which this analysis applies include the linear potential $V=igx$ and the sinusoidal potential $V=ig\sin(\alpha x)$. However, the one-turning-point analysis fails to give the full structure of the spectrum for the cubic potential $V=igx^3$, and in particular it fails to reproduce the critical points at which two real eigenvalues merge and become a complex-conjugate pair. The present paper extends the method to cases where the WKB path goes through a {\it pair} of turning points. The extended method gives an extremely accurate approximation to the spectrum of $V=igx^3$, and more generally it works for potentials of the form $V=igx^{2N+1}$. When applied to potentials with half-integral powers of $x$, the method again works well for one sign of the coupling, namely that for which the turning points lie on the first sheet in the lower-half plane.
\end{abstract}

\maketitle

\section{Introduction}
\label{intro}
The study of $PT$-symmetric Hamiltonians~\cite{BB}-\cite{AMh} has largely concentrated on eigenvalue problems in which the wave function is required to vanish at infinity in various Stokes wedges. However, some work has also been done on Sturm-Liouville problems defined on a finite interval~\cite{Uwe}, where the wave-function is required to vanish at the end points. Such problems, especially with the potential $V(x)=igx$, are relevant to a number of physical situations~\cite{Rub}-\cite{Shk}. In a previous paper~\cite{p1} we showed that for the imaginary linear potential and for potentials of the form $V(x)=ig\sin(\alpha x)$, the WKB approximation with the WKB path passing through a single turning point gives a spectacularly good picture of the spectrum, even for low energies. In particular, it correctly locates the critical points, where pairs of real eigenvalues merge and become complex conjugates of one other. However, for the potential $V(x)=igx^3$ the method failed to produce critical points, and the question of whether a different WKB path could do so was not addressed. For the $igx^3$ potential there are three turning points, and an alternative path passes through a pair of $PT$-symmetric turning points.

In the present paper we develop the formalism for such a path and show in Sec.~II that the two-turning-point WKB approximation gives a surprisingly simple secular equation. In Sec.~III we demonstrate that this approximation accurately reproduces the full structure of the spectrum for $V(x)=igx^3$. The method is also shown to give good results for higher-power potentials of the form $V(x)=igx^{2N+1}$, where $N$ is integral. When the method is applied to potentials with half-integral powers of $x$, the double-valued nature of the potential must be taken into account. We reproduce the spectrum when there exists a simple path that passes through the turning points and remains on the first Riemann sheet. In Sec.~IV we give some brief concluding remarks.

\section{WKB Calculation of Eigenvalues and Critical Points}

\label{thy}
We consider the eigenvalue equation
\begin{equation}
-\psi''(x)+V(x)\psi(x)=\lambda\psi(x)
\label{e1}
\end{equation}
defined on the interval $[-L,L]$, with Dirichlet boundary conditions $\psi(\pm L)=0$.
In what follows we take $L=1$ without loss of generality because $L=1$ can be obtained by rescaling for the monomial potentials $V(x)=-g(ix)^M$ that we consider. The equation and the boundary conditions are $PT$ symmetric, so we expect real eigenvalues when the $PT$ symmetry is unbroken; that is, when the wave functions are themselves $PT$ symmetric.  When the $PT$ symmetry is broken, the eigenvalues occur in complex-conjugate pairs.

With $\lambda=a g$ the equation becomes
\begin{equation}
\psi''(x)+gQ(x)\psi(x)=0,
\label{e2}
\end{equation}
where $Q(x)= a-V(x)$ and $a$ is considered as a fixed parameter.
For large $g$ the WKB approximation to $\psi(x)$ has the form
\begin{equation}
\label{WKB}
\psi(x) \sim \frac{1}{(gQ)^{1/4}}\left[A \exp\left(i \int_{x_0}^x ds\ \sqrt{gQ(s)}\ \right)+B \exp\left(-i \int_{x_0}^x ds\ \sqrt{gQ(s)}\ \right)\right].
\end{equation}
However, the path connecting $-1$ to $+1$ is not specified. If it goes directly from $-1$ to $+1$ without passing through a turning point, $\psi(x)$ has the same form along the path. Then, by imposing the boundary conditions $\psi(\pm 1)=0$, one obtains the standard no-turning-point approximation
\begin{equation}
\sin{I_T}=0,
\label{0TP}
\end{equation}
where $I_T\equiv\int_{-1}^1 ds\ \sqrt{gQ(s)}$. The eigenvalues are then obtained from this secular equation, which always gives real eigenvalues and predicts no degeneracies (critical points) in the spectrum.

When, as in Ref.~\cite{p1}, the path goes through a single turning point on the imaginary axis at $x_0=ib$, the wave function has different coefficients to the left and to the right of the turning point. These coefficients are determined by matching asymptotically the separate WKB approximations to the Airy functions that approximate the solution to the Schr\"odinger equation in the vicinity of the turning point. In that case the resulting secular equation is found to be
\begin{equation}
\sin{I_T}+\half e^\Delta=0,
\label{1TP}
\end{equation}
where $\Delta=2 {\rm Im} I_+$ with $I_+\equiv \int_{x_0}^1 ds\ \sqrt{gQ(s)}$. Because of $PT$ symmetry, the total integral $I_T$ can be written as $I_T=2 {\rm Re} I_+$.

It is the presence of the second term in this new secular equation that gives the interesting structure and critical points of the spectrum. When $\lambda$ is real, $\Delta$ is real and intrinsically large in magnitude. When $\Delta$ is negative, the second term is generically very small, so that the equation essentially reduces to (\ref{0TP}); when $\Delta$ is positive, the second term is generically large and cannot be balanced by the first term unless $\lambda$ becomes complex. The critical points where this change of behaviour occurs are given by $\Delta\ll 1$.

In the case of the potential $V(x)=igx^3$ there are three turning points at $x=ib(1,\omega,\omega^2)$, where $b=a^{1/3}$ and $\omega=e^{2i\pi/3}$. Numerical calculations~\cite{p1} show a series of critical points, which the one-turning-point WKB approximation fails to reproduce. We are therefore motivated to develop the formalism for a two-turning-point approximation, where the path of integration passes through the turning points $x_L=ib\omega$ and $x_R=ib\omega^2$. In this case we need three different WKB approximations to the wave function: $\psi_L(x)$ for $x$ between $-1$ and $x_L$, $\psi_M(x)$ for $x$ between $x_L$ and $x_R$, and $\psi_R(x)$ for $x$ between $x_R$ and 1. These approximations take the form
\bea
\psi_L(x)&\sim& \frac{1}{(gQ)^{1/4}}\left[L_1 \exp\left(i \int_x^{x_L} ds\ \sqrt{gQ(s)}\ \right)+L_2 \exp\left(-i \int_x^{x_L} ds\ \sqrt{gQ(s)}\ \right)\right],\cr&&\cr
\psi_M(x) &\sim& \frac{1}{(gQ)^{1/4}}\left[M_1 \exp\left(i \int_{x_L}^x ds\ \sqrt{gQ(s)}\ \right)+M_2 \exp\left(-i \int_{x_L}^x ds\ \sqrt{gQ(s)}\ \right)\right],\cr&&\cr
\psi_R(x)&\sim& \frac{1}{(gQ)^{1/4}}\left[R_1 \exp\left(i \int_{x_R}^x ds\ \sqrt{gQ(s)}\ \right)+R_2 \exp\left(-i \int_{x_R}^x ds\ \sqrt{gQ(s)}\ \right)\right].
\eea
We will not need $\psi_R(x)$ explicitly, as our tactic will be to work with $\psi_L(x)$ and $\psi_M(x)$ in the left half-plane and to enforce $PT$ symmetry of the wave function on the imaginary axis.

The first constraint on the coefficients is that $\psi_L(x)$ should vanish at $x=-1$, which gives the equation
\begin{equation}\label{C1}
\frac{L_1}{L_2}=-e^{-2iI_L},
\end{equation}
where $I_L\equiv \int_{-1}^{x_L} ds \sqrt{gQ(s)}$. The second constraint comes from requiring that the wave function be $PT$ symmetric, namely, that $\psi'(x)/\psi(x)$ be pure imaginary at an arbitrary point $x=-i\alpha$ on the imaginary axis. After some algebra this leads to the equation
\begin{equation}\label{C2}
\frac{M_1 M_2^*}{M_2 M_1^*} =e^{-2iI_M},
\end{equation}
where $I_M\equiv \int_{x_L}^{x_R} ds \sqrt{gQ(s)}$.

Finally, we match the two WKB approximations to the superposition of Airy functions
\begin{equation}
\psi_A(x)=K_1 {\rm Ai}(y)+K_2 {\rm Ai}(\omega^2 y),
\end{equation}
with $y=(x-x_L)/c$, that approximates the solution to the Schr\"odinger equation in the vicinity of $x_L$. The coefficient $c$ is given by $c=\gamma e^{-i\theta/3}$, where $\gamma=\left[b/(3ga)\right]^{1/3}$ and $\theta=5\pi/6$.
Because $g$ is large, $c$ is small. Hence, $y$ is large, and the Airy functions have two distinct asymptotic behaviours depending on the argument of $y$, namely
\bea\label{Airy}
{\rm Ai}(y)&\sim& \frac{1}{2\sqrt{\pi}y^{1/4}} e^{-\frac{2}{3}y^{3/2}}\qquad (|{\rm arg\ }y|<\pi),\cr
{\rm Ai}(y)&\sim& \frac{1}{2\sqrt{\pi}y^{1/4}}\left( e^{-\frac{2}{3}y^{3/2}}+ie^{\frac{2}{3}y^{3/2}}\right)\qquad(|{\rm arg\ }y|=\pi).
\eea
In the neighbourhood of $x_L$ the two WKB approximations can be further approximated as
\bea
\psi_L(x)&\propto&  \frac{1}{y^{1/4}}\left(L_1 e^{-\frac{2}{3} y^{3/2}}+L_2 e^{\frac{2}{3}y^{3/2}}\right),\cr&&\cr
\psi_M(x)&\propto&  \frac{1}{y^{1/4}}\left(M_1 e^{\frac{2}{3}y^{3/2}}+M_2 e^{-\frac{2}{3}y^{3/2}}\right).
\eea

\begin{figure}[h!]
\begin{center}
\includegraphics[scale=0.55]{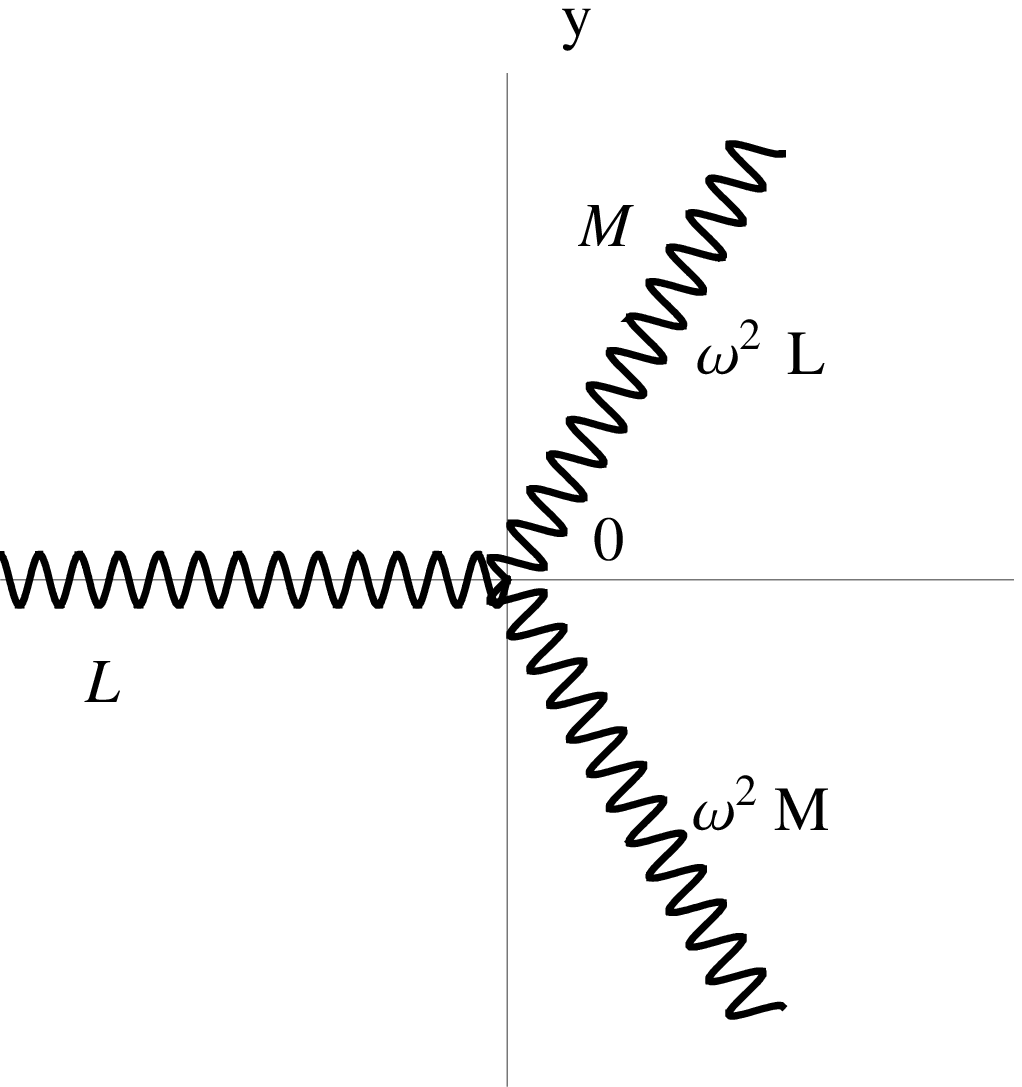}
\hspace{1.5cm}
\includegraphics[scale=0.55]{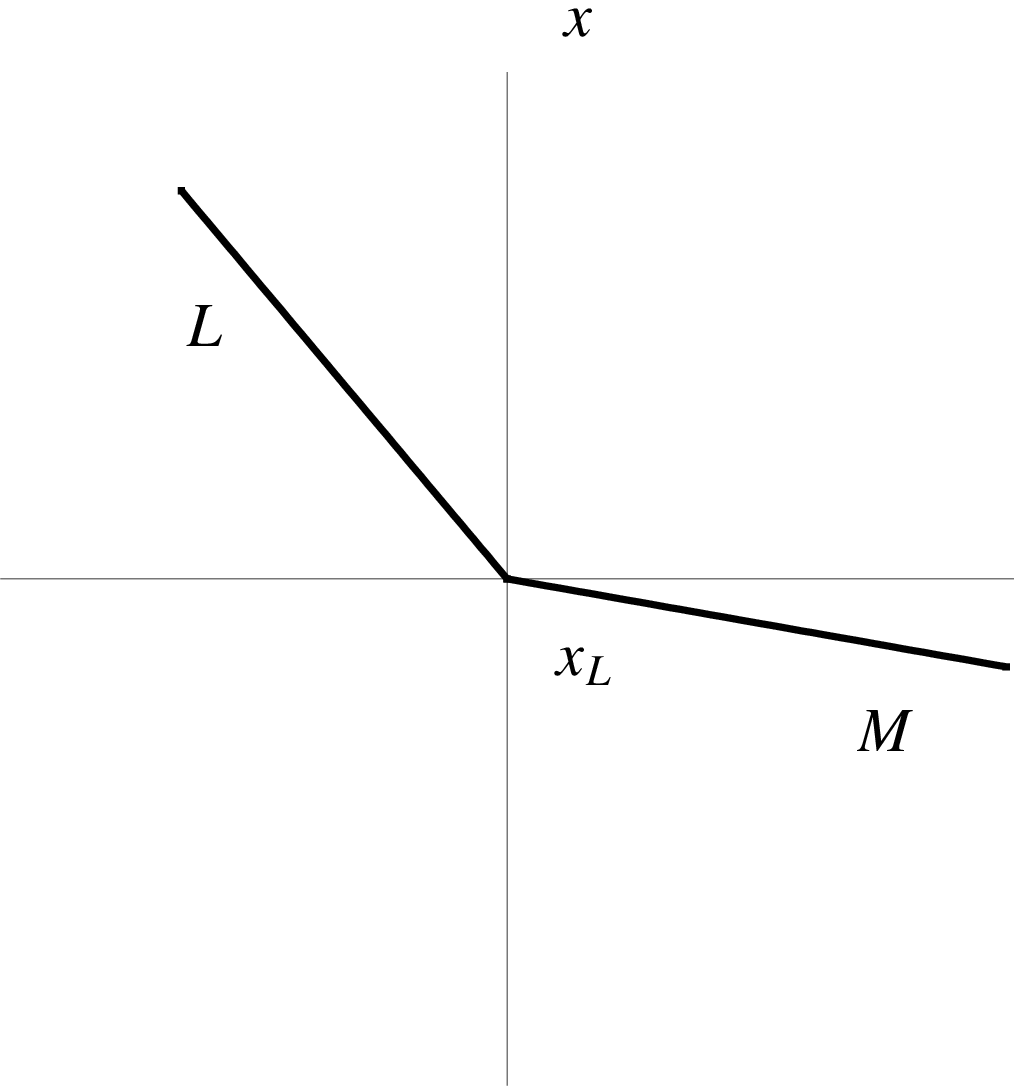}
\end{center}
\caption{Matching paths in the complex $y$-plane (left panel) and $x$-plane (right panel). }
\label{matching}
\end{figure}

In the complex-$y$ plane the matches must be made along the Stokes lines where the wave function is purely oscillatory. These Stokes lines are indicated in the left panel of Fig.~\ref{matching}. The rotation angle between $x$ and $y$ is $\theta/3$, which is approximately $\pi/3$, and the paths have to be adjusted so that $y$ and $\omega^2 y$ lie on Stokes lines. The required directions in the complex-$x$ plane are indicated in the right panel of Fig.~\ref{matching}.
For the matching to $\psi_M(x)$, both $y$ and $\omega^2 y$ lie away from the negative-$y$ axis. Thus, in this case we can use the first asymptotic approximation of (\ref{Airy}). This results in the matching
\begin{equation}
M_1 e^{\frac{2}{3}y^{3/2}}+M_2 e^{-\frac{2}{3}y^{3/2}}\propto K_1 e^{-\frac{2}{3}y^{3/2}}+K_2 e^{i\pi/6} e^{\frac{2}{3}y^{3/2}},
\end{equation}
and leads to the condition
\begin{equation}\label{C3}
\frac{M_1}{M_2}=\frac{K_2}{K_1}  e^{i\pi/6}.
\end{equation}
For the matching to $\psi_L(x)$, $\omega^2 y$  again lies away from the negative-$y$ axis, but $y$ lies along that axis. In that case we use the second approximation of (\ref{Airy}) for Ai($y$). The resulting match is
\begin{equation}
L_1 e^{-\frac{2}{3}y^{3/2}}+L_2 e^{\frac{2}{3}y^{3/2}}\propto K_1\left( e^{-\frac{2}{3}y^{3/2}}+ i e^{\frac{2}{3}y^{3/2}}\right)+K_2 e^{i\pi/6} e^{\frac{2}{3}y^{3/2}},
\end{equation}
which gives the condition
\begin{equation}\label{C4}
\frac{L_2}{L_1}=i+\frac{K_2}{K_1} e^{i\pi/6}.
\end{equation}

Combining (\ref{C1}), (\ref{C2}), (\ref{C3}) and (\ref{C4}), we obtain
\begin{equation}
e^{-2iI_M}=\frac{i+e^{2iI_L}}{-i+e^{-2iI_L^*}},
\end{equation}
which results in the surprisingly simple secular equation
\begin{equation}\label{2TP}
\sin{I_T} +e^{\Delta} \cos{I_M}=0,
\end{equation}
where $\Delta\equiv 2 {\rm Im} I_L$.

The secular equation (\ref{2TP}) is our main result. It was derived explicitly for the case $V(x)=igx^3$, but in fact it applies without change for potentials of the form $V(x)=-g(ix)^{2N+1}$ when the two relevant turning points lie in the lower-half plane. If the relevant turning points lie in the upper-half plane, the sign of $\Delta$ in (\ref{2TP}) should be reversed. When $N$ is an integer, the spectrum is the same for $g$ negative because changing the sign of $g$ is equivalent to changing the sign of $i$. Otherwise, the spectrum is asymmetric in $g$.

A few words about the role of $\Delta$ are in order. As already mentioned, it is the change in sign of $\Delta$ that produces the interesting structure in the spectrum and determines the location of the critical points. The significance of $\Delta=0$ is that when $\Delta$ vanishes, $I_L$ is purely real and there is a purely oscillatory path for the WKB wave function in going from $x=-1$ to $x=x_L$. There is a path connecting $x_L$ and $x_R$ along which the wave function is purely oscillatory, so the wave function is oscillatory along the entire path from -1 to 1. If there is no oscillatory path starting from $x=-1$ that passes through $x_L$, the WKB approximation fails to produce critical points. These considerations guide us in the choice of the appropriate turning points.
\section{Numerical calculations}
\label{num}
\begin{figure}[h!]
\begin{center}
\includegraphics[scale=0.7]{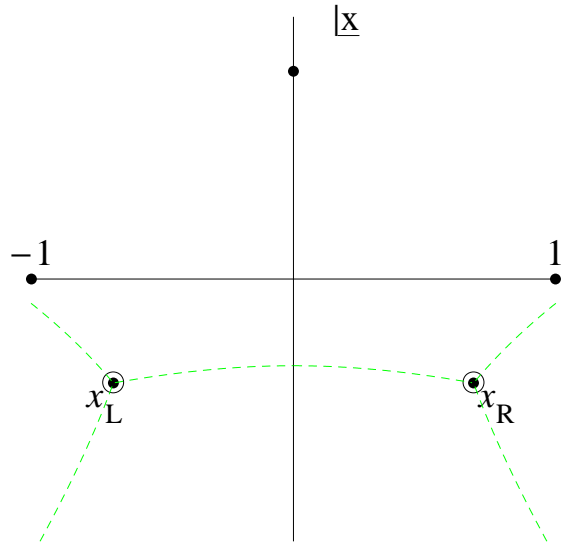}
\caption{Turning points for the potential $V=igx^3$ with $a=0.5$. The two turning points used in the WKB approximation are circled and marked $x_L$ and $x_R$. Also shown (dashed green lines) are the Stokes lines emanating from $x_L$ and $x_R$.}
\label{ix3TP}
\end{center}
\end{figure}

For $V=igx^3$ the two relevant turning points are shown circled in Fig.~\ref{ix3TP} for $a=0.5$. We also show the Stokes lines emanating from $x_L$ and $x_R$ along which the exponent in the WKB approximation is purely imaginary.  There is a purely oscillatory path between $x_L$ and $x_R$, and the continuations towards $\pm 1$ almost pass through those points. They do so when $\Delta=0$.

\begin{figure}[h!]
\begin{center}
\includegraphics[scale=1]{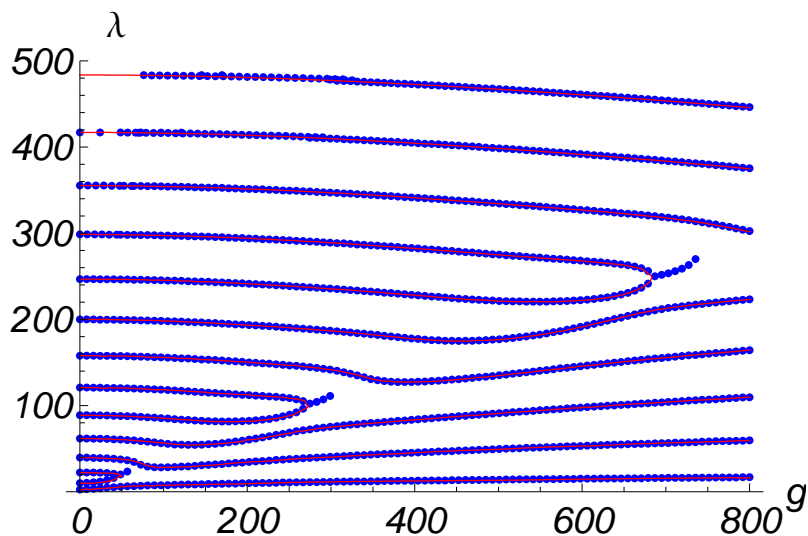}
\caption{WKB approximation (red line) for the real energy levels of the potential $V(x)=igx^3$ compared with the numerical results (blue dots). The WKB approximation and the numerical results are virtually indistinguishable. }
\label{ix3}
\end{center}
\end{figure}

Figure \ref{ix3} shows the result of (\ref{2TP}) for the real spectrum of the potential $V=igx^3$ along with the numerical results obtained by a shooting method. As can be seen, the WKB approximation reproduces the numerical result very accurately. The blue dots extending a little way beyond a critical point represent the common real part of the complex conjugate pair of eigenvalues. The spectrum for $g$ negative is not shown because it is identical to that for $g$ positive.

\begin{figure}[h!]
\begin{center}
\includegraphics[scale=0.7]{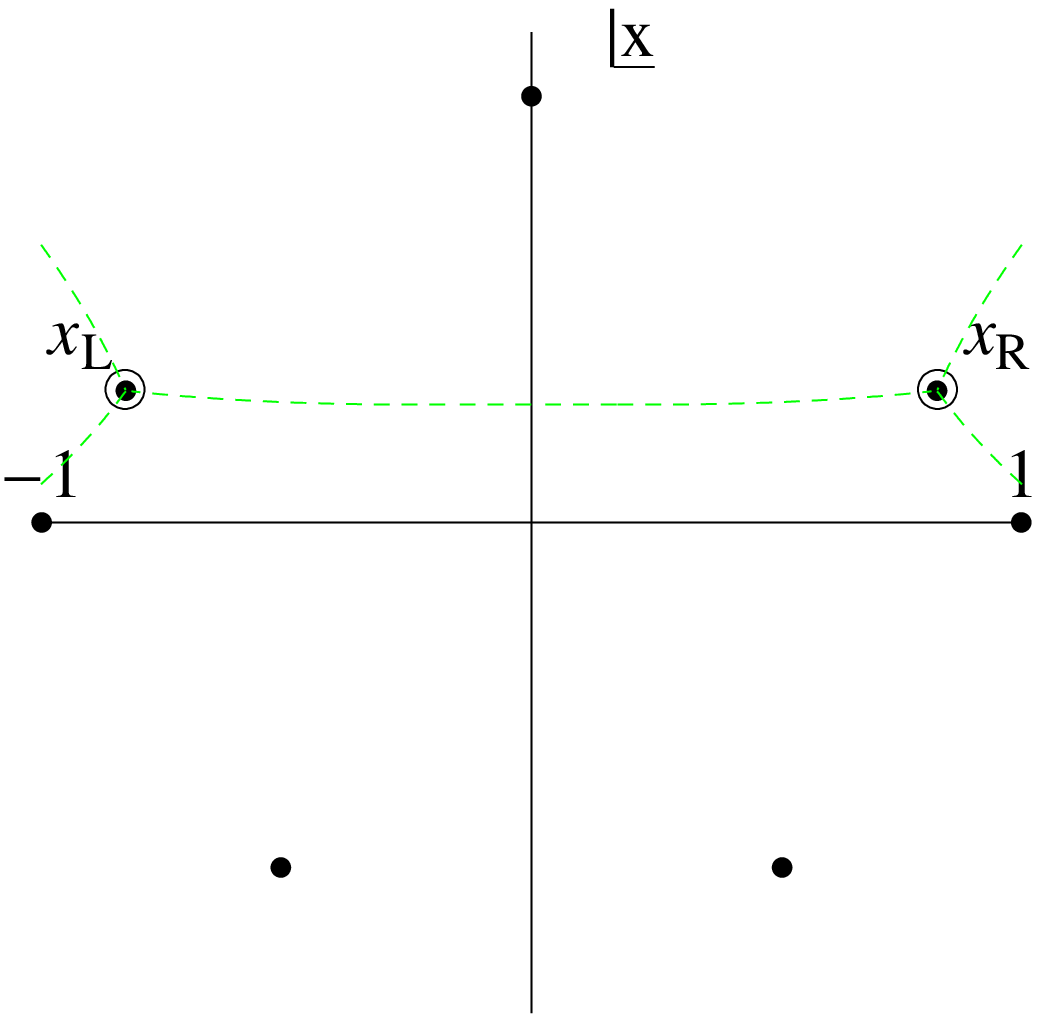}
\caption{Turning points for the potential $V=-igx^5$ with $a=0.5$. The two turning points used in the WKB approximation are circled and marked $x_L$ and $x_R$. Also shown (dashed green lines) are the Stokes lines emanating from $x_L$ and $x_R$.}
\label{ix5TP}
\end{center}
\end{figure}

\begin{figure}[h!]
\begin{center}
\includegraphics[scale=1]{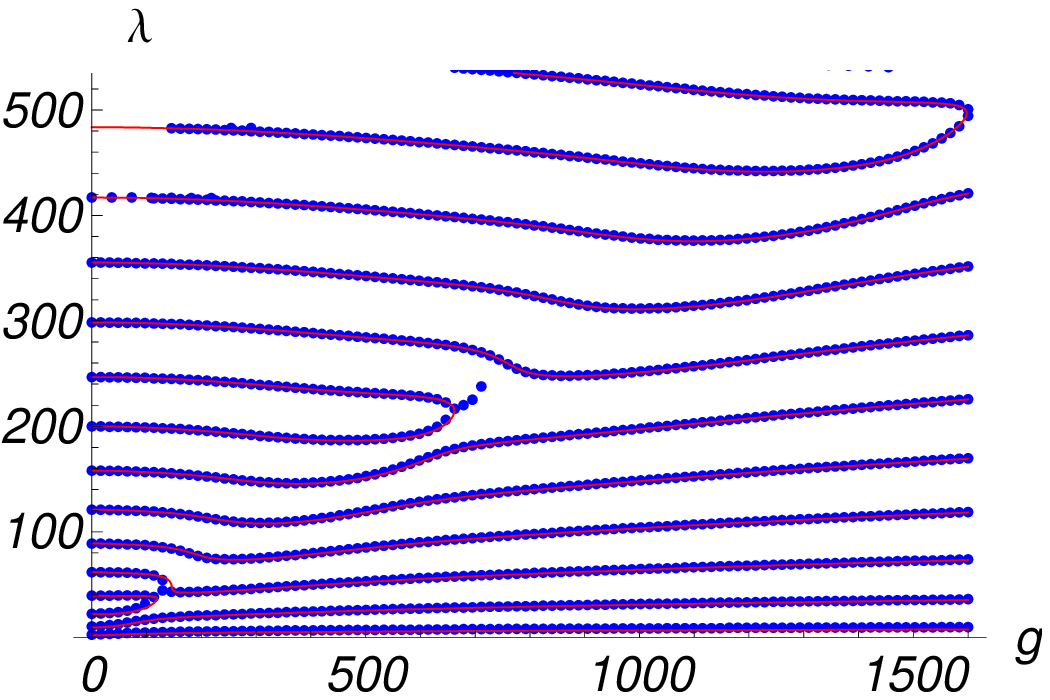}
\caption{WKB approximation (red line) for the real energy levels of the potential $V(x)=-igx^5$ compared with the numerical results (blue dots). As in Fig.~\ref{ix3}, the WKB approximation and the numerical results are virtually indistinguishable.  }
\label{ix5}
\end{center}
\end{figure}

Figures \ref{ix5TP} and \ref{ix5} show the corresponding results for the potential $V=-igx^5$. Because the relevant turning points are above the real-$x$ axis, we must now use $e^{-\Delta}$ in (\ref{2TP}). Again, the WKB approximation reproduces the numerical result extremely accurately. A general feature of the spectrum is that as the power $M$ of $x$ increases, the interesting structure occurs for higher values of $g$. This is readily understood because the problem is initially posed in the range $-1\le x \le 1$.

It seems clear that this WKB approximation will work equally well for potentials of the form $V=-(igx)^{2N+1}$, where $N$ is integral. The turning points chosen are probably those closest to the $x$-axis, but in any case they must be a pair for which the Stokes lines from $x_L$ and $x_R$ pass through $x=\pm 1$ for some value of $a$. This corresponds to $\Delta=0$ and the onset of structure in the spectrum, resulting in critical points.

We have also examined the effectiveness of the WKB approximation for half-integral $PT$-symmetric potentials. Specifically we have looked at the cases $V=-g(ix)^M$ for $M=1/2$, $3/2$ and $5/2$. Now we have to differentiate between the cases $g$ positive and $g$ negative.

\begin{figure}[h!]
\begin{center}
\includegraphics[scale=1]{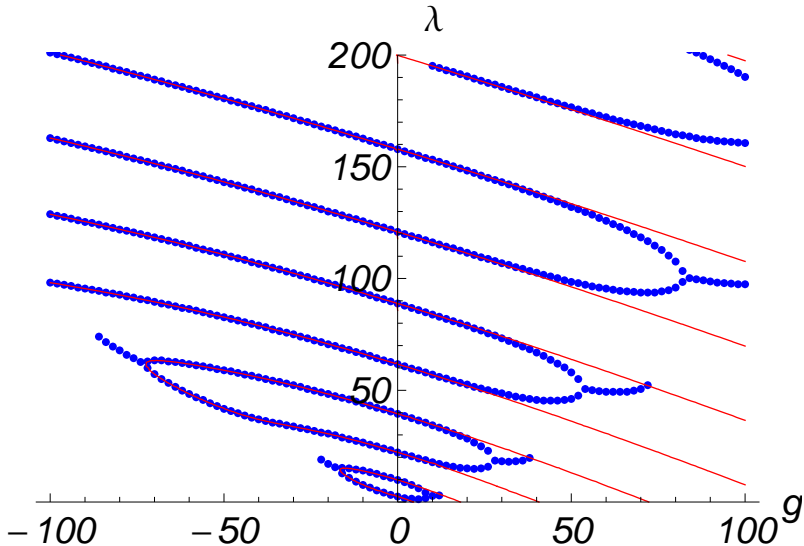}
\caption{WKB approximations (red line) for the real energy levels of the potential $V(x)=-g(ix)^\half$ compared with the numerical results (blue dots). For $g>0$ we use (\ref{0TP}), while for $g<0$ we use (\ref{1TP}) with $e^{\Delta}$ replaced by $e^{-\Delta}$. See text.}
\label{ix0p5}
\end{center}
\end{figure}

Starting with $V=-g(ix)^{1/2}$, the potential is a two-sheeted function with a cut that we take to lie along the positive imaginary axis. There is a single turning point, which is located at $x=-ia^2$, but it is on different Riemann sheets depending on the sign of $g$. For $g<0$ the turning point is on the first sheet; however, for $g>0$ it is on the second sheet. 
It is natural to take the end points $x=\pm 1$ to be on the first Riemann sheet. Then, for $g<0$ there is a straightforward WKB path through the turning point. On the other hand, for $g>0$ the path would have to pass through the cut twice in order to pass through the turning point. We do not know how to implement such a path, so for $g>0$ we simply use the straightforward no-turning-point formula of (\ref{0TP}). By its nature this latter approximation is rather smooth and does not reproduce the critical points that occur in the numerical solution. The two WKB approximations and the numerical data are presented in Fig.~\ref{ix0p5}.

\begin{figure}[h!]
\begin{center}
\includegraphics[scale=0.7]{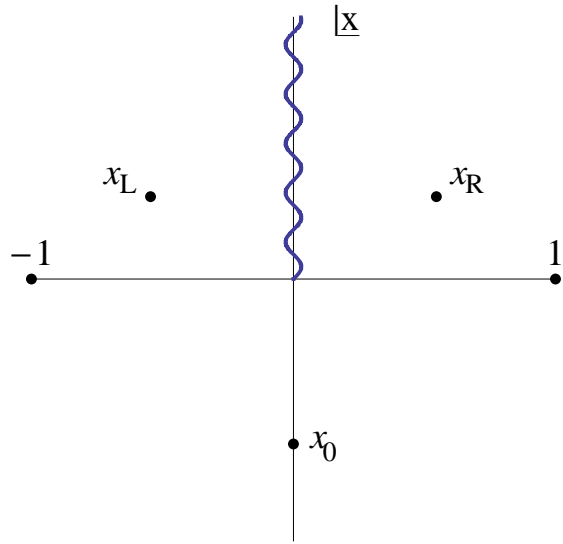}
\caption{Turning points for the potential $V(x)=-g(ix)^{3/2}$, for $a=0.7$. }
\label{tp1p5}
\end{center}
\end{figure}

Going on to $M=3/2$, the turning points are shown in Fig.~\ref{tp1p5}. There is a single turning point, marked $x_0$, at $x=-ia^{2/3}$, and a pair of turning points, $x_L$ and $x_R$, in the upper half plane. For $g<0$ the single turning point $x_0$ is on the first sheet, and the appropriate path is one going through that point, giving the WKB approximation of (\ref{1TP}). As can be seen from Fig.~8, the approximation is extremely good. For $g>0$, the single turning point is on the second sheet, while $x_L$ and $x_R$ are on the first sheet. However, there is no Stokes lines between them because of the presence of the cut. This explains why the WKB approximation of (\ref{2TP}) with $e^{\Delta}$ replaced by $e^{-\Delta}$ fails to reproduce completely the numerical structure of the spectrum.
\begin{figure}[h!]
\begin{center}
\includegraphics[scale=1]{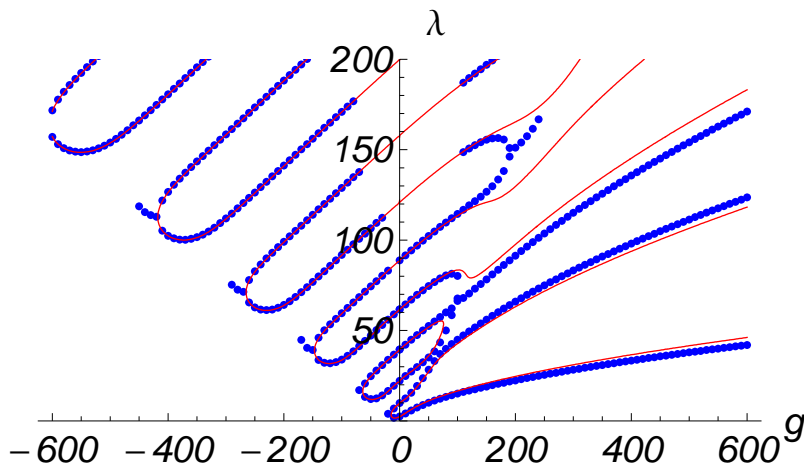}
\caption{WKB approximations (red line) for the real energy levels of the potential $V(x)=-g(ix)^{3/2}$ compared with the numerical results (blue dots). For $g<0$ we use (\ref{1TP}), while for $g>0$ we use (\ref{2TP}) with $e^{\Delta}$ replaced by $e^{-\Delta}$. This fails to reproduce the spectrum because of complications due to the cut. }
\end{center}
\label{WKBforMeq1p5}
\end{figure}
\label{}
Finally we turn to $M=5/2$. The relevant turning points are shown in Fig.~\ref{tp2p5}. For $g>0$ the turning points $x_L$ and $x_R$ are on the first sheet, and the appropriate path is one going through them, giving the extremely accurate WKB approximation of (\ref{2TP}).  For $g<0$,  the upper pair of turning points are on the first sheet, but again there is no Stokes lines between them because of the presence of the cut and consequently the WKB approximation fails to reproduce completely the numerical structure of the spectrum.
\begin{figure}[h!]
\begin{center}
\includegraphics[scale=0.7]{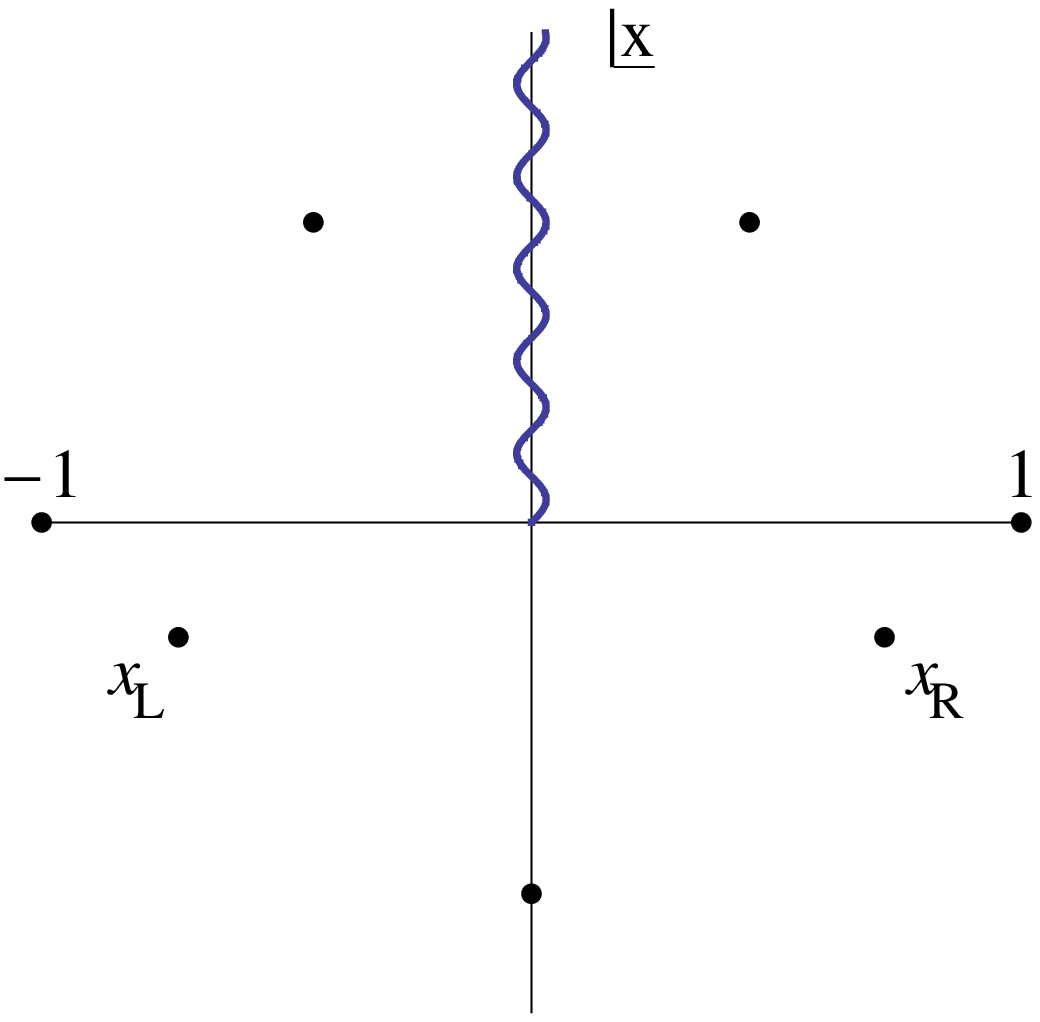}
\caption{Turning points for the potential $V(x)=-g(ix)^{5/2}$, for $a=0.7$. }
\label{tp2p5}
\end{center}
\end{figure}
\begin{figure}[h!]
\label{WKBforMeq2p5}
\begin{center}
\includegraphics[scale=1]{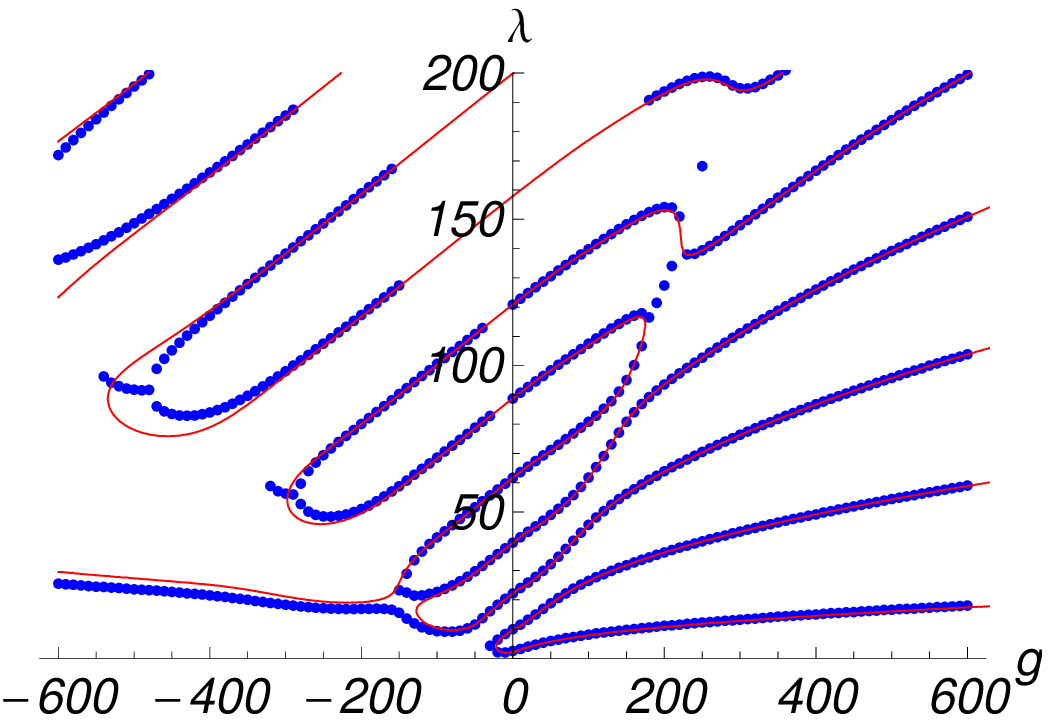}
\caption{WKB approximations (red line) for the real energy levels of the potential $V(x)=-g(ix)^{3/2}$ compared with the numerical results (blue dots). For $g>0$ we use (\ref{2TP}) with the turning points $x_L$ and $x_R$, while for $g<0$ we use the upper pair of turning points. This latter approximation fails to reproduce the spectrum because of complications due to the cut. }
\end{center}
\end{figure}
\section{Concluding remarks}
\label{comm}

We have derived a very simple secular equation for the real eigenvalues of $PT$-symmetric Sturm-Liouville problems where the potential is of the form $V=igx^{2N+1}$ with $N$ integral. This equation is derived using the WKB approximation for a path that passes through the pair of turning points lying nearest to the $x$-axis, and supplements the equation previously derived for a path passing through a single turning point. For the above potentials the spectrum is symmetric in $g$, so that it is only necessary to consider one sign of $g$. The spectra so obtained are extremely accurate and show all the interesting structure, including the critical points where two eigenvalues merge and become complex conjugates.

We also considered the extension of the secular equation to potentials with half-integral powers. In those cases the situation is complicated by the presence of a cut, and a straightforward application is only possible for the particular sign of $g$ where the relevant critical point or points lie in the lower-half plane. Here the results are again extremely accurate. For the other sign of $g$ the method fails to reproduce the complete structure of the spectrum, and an extension of the method is needed to deal with the situation when either the turning points are on a different sheet from the end-points $\pm 1$ or the path between them is impeded by the cut.

As mentioned in the Introduction, this problem with the linear potential $V=igx$ occurs in a number of physical situations, and one may speculate that the higher-power potentials considered in this paper might also have some physical relevance. The most likely application is in the context of hydrodynamics, discussed in Ref.~\cite{Shk}, where the potential $V(x)$ is associated with the velocity profile $q(x)$ of the fluid.

\begin{acknowledgments}
CMB is supported by the U.K.~Leverhulme Foundation and by the U.S.~Department of Energy.
\end{acknowledgments}

\end{document}